\documentclass[aps,prd,twocolumn,superscriptaddress]{revtex4}
\usepackage{graphicx}
\usepackage{amsmath}
\usepackage{epsfig}
\usepackage{amsmath,amsfonts,amssymb}
\usepackage{times}
\usepackage{epsfig}
\usepackage{epstopdf}
\usepackage[colorlinks]{hyperref}

\usepackage{subcaption}
\usepackage[export]{adjustbox}

\usepackage{color}

\begin{document}
\baselineskip=12pt
\def\black{\textcolor{black}}
\def\red{\textcolor{black}}
\def\blue{\textcolor{blue}}
\def\green{\textcolor{black}}
\def\be{\begin{equation}}
\def\ee{\end{equation}}
\def\bea{\begin{eqnarray}}
\def\eea{\end{eqnarray}}
\def\orc{\Omega_{r_c}}
\def\om{\Omega_{\text{m}}}
\def\E{{\rm e}}
\def\bearst{\begin{eqnarray*}}
\def\eearst{\end{eqnarray*}}
\def\peleven{\parbox{11cm}}
\def\peffec{\peight{\bearst\eearst}\hfill\peleven}
\def\pspace{\peight{\bearst\eearst}\hfill}
\def\ptwelve{\parbox{12cm}}
\def\peight{\parbox{8mm}}

\newcommand{\nk}[1]{\textcolor{red}{[{\bf Nima}: #1]}}

\newcommand{\hm}[1]{\textcolor{blue}{[{\bf Hossein}: #1]}}


\title{ CMB lensing in a modified $\Lambda$CDM model in light of the $H_0$ tension}

\author{Hossein Moshafi}
\email{moshafi@ipm.ir}
\address{School of Astronomy, Institute for Research in Fundamental Sciences (IPM), P.~O.~Box 19395-5531, Tehran, Iran}
\author{Shant Baghram}
\email{baghram@sharif.edu}
\address{Department of Physics, Sharif University of Technology, P.~O.~Box 11155-9161, Tehran, Iran}
\author{Nima Khosravi}
\email{n-khosravi@sbu.ac.ir}
\address{Department of Physics, Shahid Beheshti University, 1983969411,  Tehran, Iran}
\address{School of Physics, Institute for Research in Fundamental Sciences (IPM), P.~O.~Box 19395-5531, Tehran, Iran}

\begin{abstract}
The observed discrepancy of the Hubble parameter measurements in the local universe with the cosmic microwave background (CMB) data may indicate a new physics. It is vital to test the alternative models that reconcile the Hubble tension with other cosmological observations in this direction. The CMB lensing is a crucial observation that relates the early universe perturbations to the matter's late-time distribution. In this work, we study the prediction of the \"u$\Lambda$CDM as a solution for $H_0$ tension for CMB lensing and the low- and high-$\ell$'s  temperature (TT) power spectrum internal inconsistency. We show that this model relaxes the low- and high-$\ell$'s TT mild inconsistency and the CMB lensing tensions simultaneously. Accordingly, \"u$\Lambda$CDM having the same number of free parameters as $\Lambda$CDM with lensing amplitude $A_L$ added, has a better fit with $\Delta \chi^2=-3.3$.

PACS numbers:
04.50.+h, 95.36.+x, 98.80.-k
\end{abstract}

\maketitle

\section{Introduction}
The recent observations of local standard candles, which leads to a more precise measurement of the cosmos' expansion rate and determination of the Hubble constant $H_0$, introduce a new challenge to cosmology's standard paradigm. The challenge is the apparent discrepancy of the local measurements of $H_0$ \cite{Riess:2016jrr,Riess:2018uxu,Riess:2019cxk} with the value obtained from cosmic microwave background (CMB) observations \cite{Aghanim:2018eyx}. There are some other anomalies between local datasets and CMB if the standard model of cosmology $\Lambda$CDM is assumed. One of the famous ones is the amount of matter content $\sigma_8$ measured locally \cite{Heymans:2020gsg} in comparison to CMB \cite{Aghanim:2018eyx}. Besides, an internal inconsistency is reported in Planck results \cite{Aghanim:2018eyx} in measurements of CMB lensing amplitude, $A_L$, which can be related to low/high-$\ell$ inconsistency. These issues, if not systematic errors, trigger many interests in cosmological model building.\\
The idea that a new physics is needed to explain this discrepancy ($H_0$ tension) opens a vast arena for model building and phenomenological predictions.
The ideas span a vast range starting from dark sector interactions \cite{Kumar:2016zpg, Xia:2016vnp, DiValentino:2017iww,Kumar:2017dnp,Gomez-Valent:2020mqn,Lucca:2020zjb,vandeBruck:2017idm,Yang:2018euj,Yang:2018uae,Yang:2019uzo,Martinelli:2019dau,DiValentino:2019ffd,DiValentino:2019jae,Pettorino:2013oxa,Yang:2020uga,Yang:2019uog,Yang:2018ubt}, early dark energy models which modify sound horizon \cite{Poulin:2018cxd,Karwal:2016vyq,Pettorino:2013ia}, phase transition in dark energy \cite{Banihashemi:2018has,Banihashemi:2018oxo,Farhang:2020sij,Li:2019yem,Pan:2019hac,Li:2020ybr} to the modified gravity models \cite{Nunes:2018xbm,Ballardini:2020iws,Braglia:2020auw}.
The Hubble constant tension, on the one hand, and the proposed models to solve this tension, on the other hand, remind us that the unknown nature of dark matter and dark energy is at the heart of the problem. In this direction, the long-standing question of cosmological constant and plausible gravity models leads to a class of solutions known as ensemble theories of gravity \cite{Khosravi:2016kfb,Khosravi:2017aqq}. This idea proposes a density-dependent transition of the law of gravity. In this direction a cosmological model is introduced known as \"u$\Lambda$CDM in \cite{Khosravi:2017hfi}. We showed \"u$\Lambda$CDM can address $H_0$ tension.\\

In this work, we want to check if \"u$\Lambda$CDM can address the CMB lensing inconsistency in addition to the Hubble tension. While having local $H_0$ measurements allows us to check the background of our model, the CMB lensing checks \"u$\Lambda$CDM at the level of its perturbations. Another reason for this choice is that the lensing in cosmological scales from the CMB to late-time lenses such as cosmic shear observations are promising tools to detect the deviations from the standard model of cosmology \cite{Hassani:2015zat, Fard:2017oex}. We also study the effect of the integrated Sachs-Wolfe (ISW) in this context. The ISW is related to the change of potential from the last scattering to the observer, and it is the other gravitational secondary effect on the CMB besides the lensing effect.\\
The structure of this work is as follows: In Sec.\ref{Sec2}, we review the ensemble theory of gravity and the \"u$\Lambda$CDM model. In Sec.\ref{Sec3}, we study the lensing and ISW of the \"u$\Lambda$CDM. In Sec.\ref{Sec4}, we discuss the data and methodology. In Sec.\ref{Sec5}, we show the results, and in Sec.\ref{Sec6}, we conclude, and we have our future remarks.
\section{Ensemble Average Theory of Gravity to \"u$\Lambda$CDM}
\label{Sec2}
The ensemble average theory of gravity \cite{Khosravi:2016kfb,Khosravi:2017aqq} suggests that the gravity model works in the Universe are an ensemble average of theoretically possible models of gravity. This idea is formulated in the Lagrangian formalism via the relation below:
\begin{equation}
{\cal{L}} = (\sum_{i=1}^{N}{\cal{L}}_i e^{-\beta {\cal{L}}_i})  / \sum_{i=1}^{N} e^{-\beta {\cal{L}}_i},
\end{equation}
where ${\cal{L}}_i$ are theoretically possible Lagrangians and $\beta$ is a free parameter  of the model. A possible and simple derivation of a model from the ensemble theory idea is the 
the \"uber-gravity gravity model which used the power law terms in the Ricci scalar as a plausible candidate of $f(R)$ gravity models which make an independent basis for model space such as 
\begin{equation}
{\cal{L}}_{\ddot{u}ber} = (\sum_{i=1}^{N}( \bar{R}^n-2\Lambda) e^{-\beta( \bar{R}^n-2\Lambda)})  / \sum_{i=1}^{N}  e^{-\beta( \bar{R}^n-2\Lambda)},
\end{equation}
where $\bar{R}=R/R_0$; in this case $R_0$ is a free parameter of the model. We call it \"u$\Lambda$CDM. This cosmological model is an extension of the standard model. 
This leads to a simple model for the gravity as \cite{Khosravi:2017hfi}
\begin{eqnarray} \label{uber-cosmology}
\text{Gravity} \simeq
\begin{cases}
\text{$R=R_0$} & \rho< \rho_{{\rm \ddot{u}}} \\
\text{$\Lambda$CDM} & \rho > \rho_{{\rm \ddot{u}}},
\end{cases}
\end{eqnarray}
where $\rho_{{\rm \ddot{u}}}$ is the critical density in which the transition occurred.
In the case, if the density is higher than the critical value which $\rho>\rho_{\ddot{u}}$, we recover the standard model with the cosmological constant and in the regime of $\rho < \rho_{\ddot{u}}$, we have constant Ricci scalar. The transition occurs in $z_{\oplus}$, where in two regimes the background evolution will be
\begin{equation}
E^2(z) = \Omega_m(1+z)^3 + \Omega_\Lambda,  ~~~~ z>z_{\oplus}
\end{equation} 
where $E(z)=H/H_0$ is the normalized Hubble parameter:
\begin{equation}
E^2(z) = \frac{1}{2} \bar{R}_0+(1-\frac{1}{2}\bar{R}_0)(1+z)^4,  ~~~~ z<z_{\oplus}
\end{equation}
where $ \bar{R}_0= R_0/6H_0^2$. The continuity condition for $E(z)$ and $dE/dz$ imposes that we have only one more free parameter $z_{\oplus}$ in comparison to $\Lambda$CDM.\\
In order to study the perturbation theory, the \"u$\Lambda$CDM behavior is exactly derived form the action below:
\begin{equation}
S=\frac{1}{16\pi G} \int d^4x \sqrt{-g} [ \xi(R-R_0)-\lambda] + {\cal{L}}_m,
\end{equation}
where $g$ is the determinant of the metric $g_{\mu\nu}$, and $R_0$ is a constant free parameter of the model. $\xi$ is the Lagrange multiplier, which ensures that after the transition redshift the Ricci scalar is constant and it is equal to $R=R_0$.  For $z>z_{\oplus}$, we want to recover the standard $\Lambda$CDM model, so in this  era $\xi=1$ and $\lambda=2\Lambda-R_0$, where $\Lambda$ is the cosmological constant in the $\Lambda$CDM.
Accordingly, this action will give us the equation of the motion and the trace of the field equation governing the dynamics of the field $\xi$:
\begin{equation} \label{eq:xi-trace}
\xi\, R_0=8\pi\, G\,T - 2\,\lambda - 3\, \Box \xi,
\end{equation}
where $T=g_{\mu\nu}T^{\mu\nu}$ is the trace of energy-momentum tensor. Now we can use the above Lagrangian to study the perturbation theory of our model after the transition in \"u$\Lambda$CDM i.e. $z<z_\oplus$.  We define the Bardeen gravitational potentials $\Psi(\vec{x},\eta)$ and $\Phi(\vec{x},\eta)$ in the perturbed Friedmann-Robertson-Walker metric as
\begin{equation}
	ds^2=a^2(\eta)[-(1+2\Psi(\vec{x},\eta))d\eta^2 + (1-2\Phi(\vec{x},\eta))d\chi^2],
\end{equation}
where $\eta$ is the conformal time. Now we perturb the quantities in the theory as below up to first order (with superscript $1$ for $\xi$ ):
\begin{eqnarray}
R \!\!&=&\!\! R_0 \,, \label{pertR} \\
\xi \!\!&=&\!\!\xi^0(\eta)+ \xi^1({\vec x},\eta)\,, \label{pertxi} \\
T_{\mu\nu}\!\!&=&\!\!\bar{T}_{\mu\nu}(\eta)+\delta T_{\mu\nu}({\vec x},\eta)\,, \label{pertTab}\\
T\!\!&=&\!\!\bar{T}(\eta)+\delta T({\vec x},\eta). \label{pertT}
\end{eqnarray}
The background equations of motion are
\begin{eqnarray}
\frac{R_0 a^2}{3}\!\!&=&\!\! 2\frac{ a ''}{a} \,,\label{0pertR}\\
\frac{R_0 a^2}{3}\!\!&=&\!\! \frac{8\pi G \bar{T} a^2}{3\xi^0}+2 {\cal{H}}\frac{{\xi^0}'}{\xi^0} + \frac{{\xi^0}''}{\xi^0} -\frac{2}{3}a^2\lambda \,, \label{0perttrace} \\
\frac{R_0 a^2}{3}\!\!&=&\!\! \frac{16\pi G \bar{T}^0_0 a^2}{3\xi^0}+2{\cal{H}}({\cal{H}}+\frac{{{\xi}^0}'}{\xi^0}) -\frac{\lambda a^2}{3\xi^{(0)}}\,, \label{0pertG00}
\end{eqnarray}
where ${\cal{H}}={a}'/ a$ and $'$ is the derivative with respect to conformal time $\eta$. \\
The equations of motion at linear order are:
\begin{widetext}
\begin{eqnarray}
\nabla^2(\Psi-2\Phi)\!\!&=&\!\! -\Psi R_0 a^2-3{\cal{H}}(3{\Phi}'+{\Psi}')-3{\Phi}''\,,\label{1pertR} \\
\frac{R_0a^2}{3}\xi^1\!\!&=&\!\!\frac{8\pi G a^2}{3}\delta T-2{\cal{H}}(2\Psi{{\xi}^0}'-{{\xi}^1}')-\Box \xi^1-{{\xi}^0}'(3{\Phi}'+{\Psi}')-2\Phi {{\xi}^0}''\,, \label{1perttrace}\\
\frac{R_0a^2}{3}\xi^1\!\!\!&=&\!\!\!\frac{16\pi Ga^2}{3}\delta T_0^0\!+\!3\xi^1{\cal{H}}^2\!+\!\frac{4}{3}\xi^0\nabla^2\Phi\!-\!{\cal{H}}({\cal{H}}\Psi\!+\!{{\Phi}}')\!-\!2{\cal{H}}(2\Phi{{\xi}^0}'-{{\xi}^1}')\!-\! \frac{2}{3}\nabla^2\xi^1\!.\label{1pertG00}
\end{eqnarray}
The $\{0i\}$ component of the equation of motion is
\begin{equation}
2\xi^0({\cal{H}}\Psi_{,i}+{\Phi}'_{,i})=8\pi G \delta T_{0i}-{\cal{H}}\xi^1_{,i}-\Psi_{,i}{{\xi}^0}'+{{\xi}^1}'_{,i}\,.
\end{equation}
\end{widetext}
Using the equation of motions in the linear scalar perturbation  we can write the Newtonian potential $\Psi$  and lensing potential $\phi_L$ in the quasistatic regime ($\nabla^2 \gg {\cal H}^2$) \cite{Khosravi:2017hfi}:
\begin{eqnarray}\label{pert}
	\nabla^2 \Psi &=&\frac{16\pi G a^2}{3\xi(z)}\delta\rho, \\ \nonumber
	\phi_L &=& \frac {\Phi + \Psi}{2} = \frac{3}{4}\Psi.
\end{eqnarray}
The continuity, Euler equation and the evolution of dark matter density contrast (growth function ) are given accordingly
\begin{equation}
{\delta}' = -\theta + 3{\Phi}'
\end{equation}
\begin{equation}
{\theta}'+ {\cal{H}}= -\nabla^2\Psi
\end{equation}
\begin{equation} \label{eq:delta}
{\delta}'' + {\cal{H}}{\delta}' - \frac{16\pi G a^{-1}}{3\xi^0(\eta)} \bar{\rho}\delta=0.
\end{equation}
 It is worth mentioning that we modify the Poisson equation by replacing $8\pi G \rightarrow 16 \pi G / (3 \xi(z))$ after the transition. On the other hand,  we have $\Phi=\Psi$ in general relativity which is not always the case in modified gravity models \cite{Baghram:2010mc}. It is the case in our model, and we change the gravitational lensing potential by replacing the $\phi_L=\Psi \rightarrow \phi_L=3\Psi/4$  after the transition. However, an important point to note here is that the lensing potential is continuous in transition redshift. This is because the $\Psi$ is $4/3$ times  the $\Lambda$CDM model. This is a crucial point, which enables us to calculate the ISW effect in the upcoming section.
 Note that for $z>z_\oplus$ we use the standard perturbation theory for the $\Lambda$CDM model and match them to the ones coming from \"u$\Lambda$CDM (\ref{pert}) at $z=z_\oplus$ and $z<z_\oplus$; we use the \"u$\Lambda$CDM perturbation theory.


\section{Planck internal (in)consistencies in \"u$\Lambda$CDM}
\label{Sec3}
In \cite{Aghanim:2018eyx},  they check the internal consistency of the CMB dataset. In this direction,  they report two (mild) inconsistencies: the low- and high-$\ell$'s  temperature (TT) power spectrums are in $\sim 2\sigma$ tension and the amplitude of lensing $A_L$ is higher than unity again around $2\sigma$. In this section, we are studying these two (in)consistencies in the context of \"u$\Lambda$CDM. We check our results against the higher $H_0$ value reported by Riess {\it{et al.}} \cite{Riess:2019cxk} to see if \"u$\Lambda$CDM can address internal inconsistencies and $H_0$ tension simultaneously. We also check ISW in our model with galaxy cross-correlation. The ISW effect is also related to the change of the gravitational potential.
\subsection{Low- and high-$\ell$'s}
\begin{figure}
	\includegraphics[width=\columnwidth]{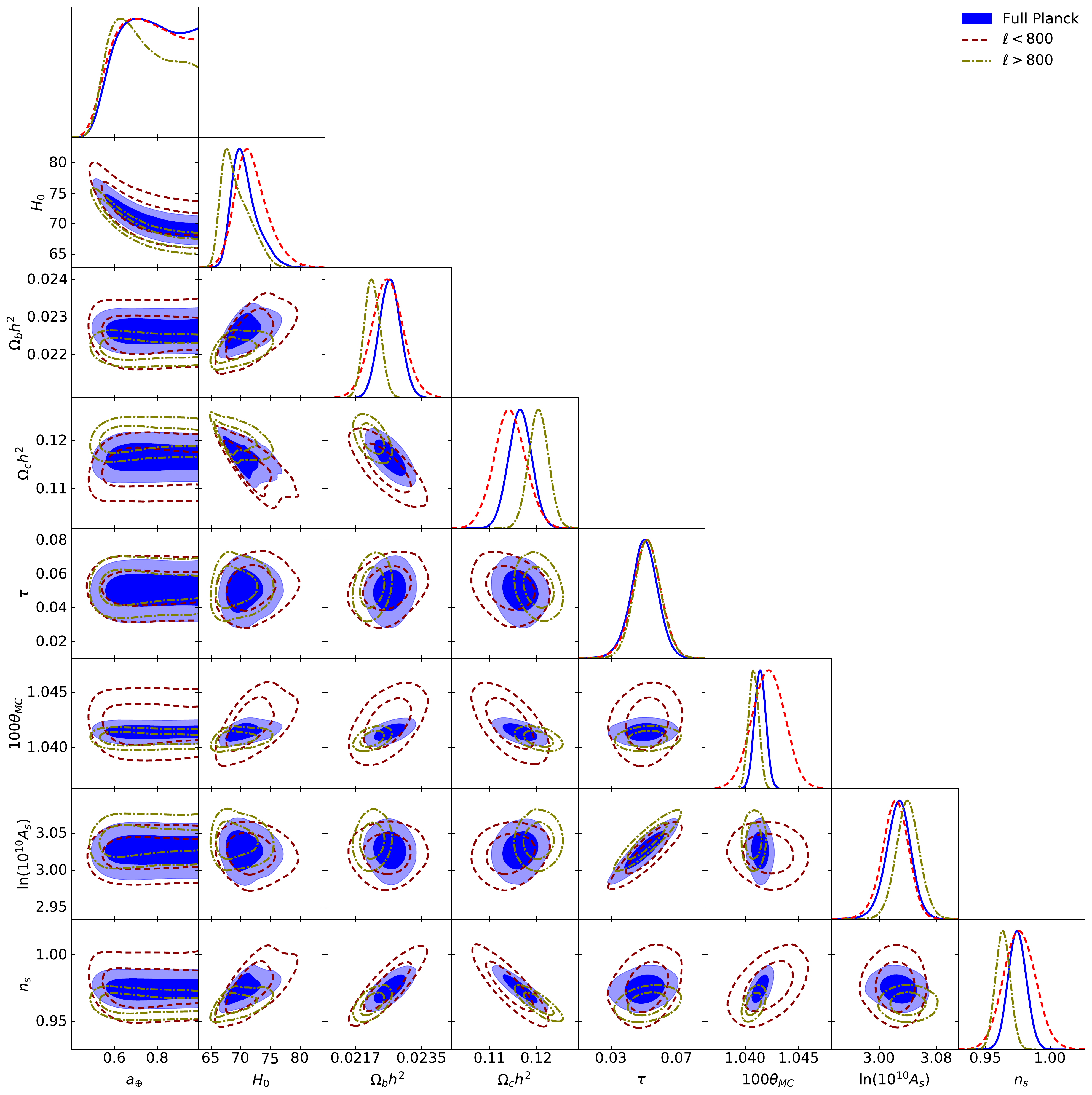}[t!]
	\captionsetup{justification=raggedright,singlelinecheck=false}
	\caption{ The contour plot of  \"u$\Lambda$CDM parameters is represented for high-$\ell$'s with green dash-dotted lines and   low-$\ell$'s with red dashed lines. The blue contour plots show the $1\sigma$ and $2\sigma$ constraints with full Planck TT data. } \label{fig1}
\end{figure} 
\begin{table*}[tb]
	\begin{center}
		\resizebox{0.5 \textwidth}{!}{
			\begin{tabular}{ c  ||c||c  }
				\hline
				\hline
				&   Planck &Planck  \\ \hline
				Parameters & $ 2< \ell < 800$ & $ 800 <\ell < 2500$  \\ \hline
				
				$\Omega_b h^2$ &  $0.02256\pm 0.00043$ & $0.02215\pm 0.00022$ \\
				$\Omega_c h^2$ &  $0.1143\pm 0.0034$ & $0.1204\pm 0.0021$ \\
				
				$\Omega_m$  &  $0.268\pm 0.025$& $0.300^{+0.026}_{-0.019}$ \\
				
				$H_0$ &  $71.9^{+2.1}_{-3.1}$ & $69.2^{+1.5}_{-2.9} $ \\
				
				$a_\oplus$ &  $0.75\pm 0.13$ & $0.73^{+0.11}_{-0.18}$  \\
				
				$S_8$ &   $0.800\pm 0.049$ &  $0.888\pm 0.036$ \\
				$100\theta_{MC} $  & $1.0422\pm 0.0015$ & $1.04080\pm 0.00047$ \\
				${\rm{ln}}(10^{10} A_s) $  &  $3.022^{+0.019}_{-0.017}$ & $3.040\pm 0.017$ \\
				$n_s $ & $0.9745\pm 0.0070$  & $0.9633\pm 0.0057$ \\
				$\tau$ &  $0.0515\pm 0.0089$ & $0.0521\pm 0.0082$ \\
				
				\hline
				\hline
				
			\end{tabular}
		}
	\end{center}
	\captionsetup{justification=raggedright,singlelinecheck=false}
	\caption{$68\%$ C.L. constraints for the \"u$\Lambda$CDM , for Planck data in two ranges $\ell < 800$ and $\ell > 800$. }
	\label{table1}
\end{table*}
We check our model against the low- and high-$\ell$ TT power spectrum separately and we show the results in Fig.\ref{fig1} and Table\ref{table1}. The contour plots show no inconsistencies between low- and high-$\ell$'s in the \"u$\Lambda$CDM framework. This result allows us to add them up together and use the full CMB TT power spectrum in our upcoming analysis.

\subsection{Lensing in \"u$\Lambda$CDM}
\begin{figure}
\includegraphics[width=\columnwidth]{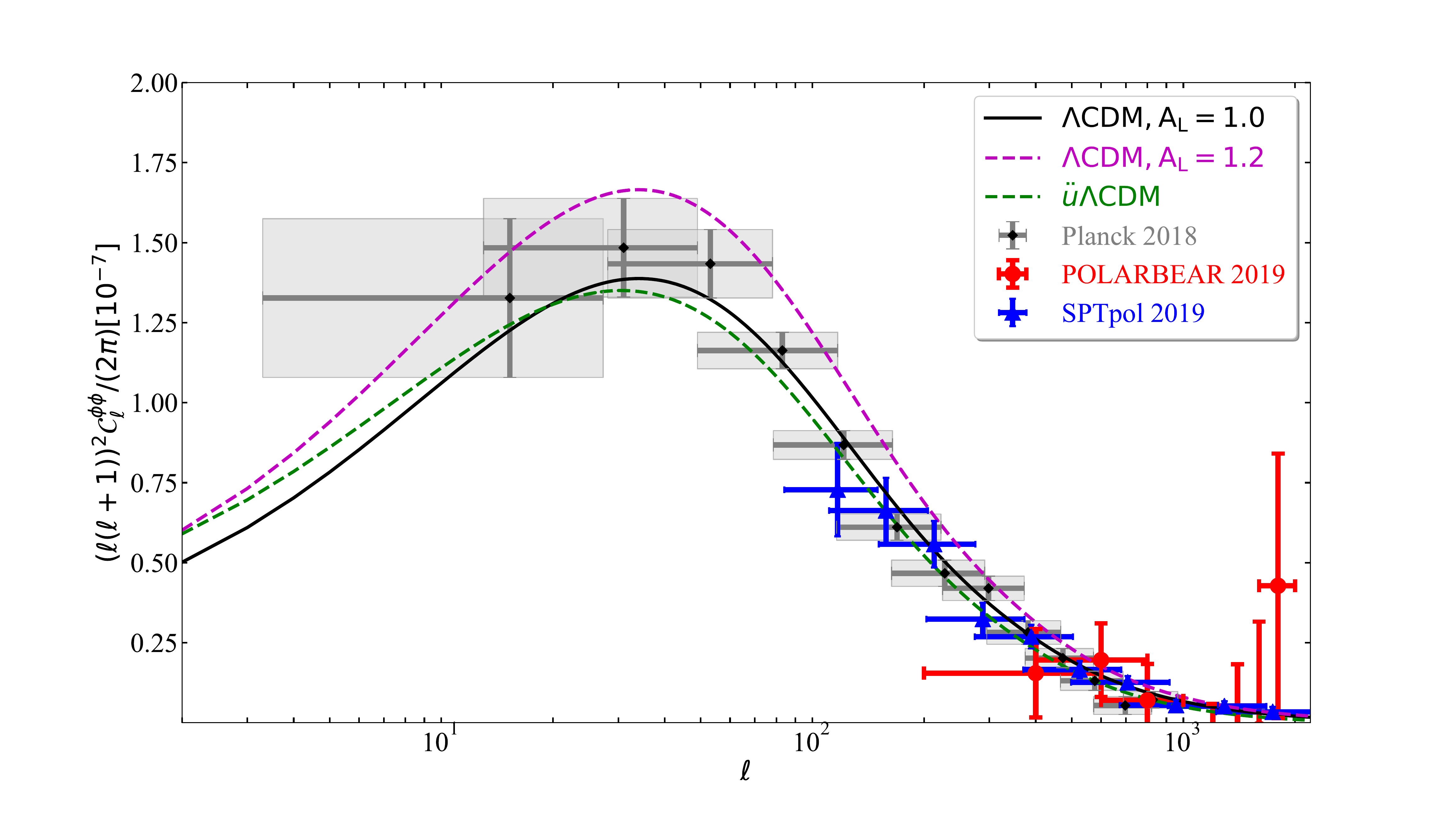}
\captionsetup{justification=raggedright,singlelinecheck=false}
\caption{The angular power spectrum of the CMB lensing is plotted for \"u$\Lambda$CDM, $\Lambda$CDM with and without free $A_L$. It is obvious that \"u$\Lambda$CDM mimics $\Lambda$CDM with $A_L=1$ without any inconsistency with the data points. Note that the temperature power spectrum prefers $\Lambda$CDM with $A_L=1.2$ which is not consistent with CMB lensing power spectrum.}\label{fig:cllens}
\end{figure}
In this section, we will study the CMB lensing in the context of \"u$\Lambda$CDM. Due to the Eq.(\ref{pert}), the lensing potential in \"u$\Lambda$CDM named as $\phi^{\ddot{u}}$ is as
\begin{equation}
\phi^{\ddot{u}} (\hat{n}) = -2 \int_{0}^{\chi_*}d\chi \left(\frac{\chi_* - \chi}{\chi_* \chi} \right) \phi_L,
\end{equation}
where $\hat{n}$ is direction of the observation, and $\chi_*$ is the comoving distance to the last scattering. Note that in the standard $\Lambda$CDM case $\phi_L=\Psi$ and $\Psi$ are obtained from the standard Poisson equation $\nabla^2 \Psi =4\pi G a^2\delta\rho$ \cite{Lewis:2006fu}.
The angular power spectrum of the lensing is obtained as
\begin{eqnarray} \label{eq:cmblensing}
\ell^4C_{\ell}^{\phi\phi}&=&18\Omega_m^2H_0^4\int_{0}^{\chi_*} d\chi \chi^2 \left(\frac{\chi_* - \chi}{\chi_*\chi} \right)^2 P^{\ddot{u}}_{m} (\frac{\ell}{\chi}) \\ \nonumber
&\times& \left[ \frac{D ^{\ddot{u}}(z)(1+z)}{D ^{\ddot{u}}(z=0)\xi(z)} \right]^2,
\end{eqnarray}
where $P^{\ddot{u}}_{m}(\frac{\ell}{\chi})$ is the matter power spectrum of \"u$\Lambda$CDM in present time and $D ^{\ddot{u}}(z)$ is the growth function of the model obtained from Eq.(\ref{eq:delta}).
We should note that the \"u$\Lambda$CDM power spectrum, growth function, and lensing potential are the same as the $\Lambda$CDM model for $z>z_{\oplus}$.
This means that for $z>z_{\oplus}$ we have $\xi(z)\rightarrow 1$, $\Phi_L\rightarrow \Psi$, $D ^{\ddot{u}}(z) \rightarrow D(z)$ and $P^{\ddot{u}}_{m} \rightarrow P_m$, where $D(z)$ and $P_m(z)$ are $\Lambda$CDM growth function and power spectrum. Accordingly, the integral in Eq.(\ref{eq:cmblensing}) is a sum of two integrals one from zero to comoving distance related to transition redshift $(0,\chi(z_{\oplus}))$ and then the other integral from $(\chi(z_{\oplus}), \chi_*)$. In first one we use the \"u$\Lambda$CDM lensing potential which leads to corresponding power spectrum and growth function and in the second part of the integral we have the $\Lambda$CDM lensing potential, integrated in line of sight.
In Fig. \ref{fig:cllens}, we plot the angular power spectrum of lensing for both standard model and \"u$\Lambda$CDM. The data points are from Planck 2018 \cite{Aghanim:2018oex}, Polarbear 2019 and SPTpol2019 \cite{Bianchini:2019vxp}.
\subsection{ ISW in \"u$\Lambda$CDM }
The integrated Sachs Wolfe (ISW) effect introduces secondary anisotropy on the CMB. ISW is the imprint of the dynamical gravitational potentials on the CMB  photons. 
The ISW effect in \"u$\Lambda$CDM is as below:
\be
\frac{\Delta T}{T}|_{ISW} =2 \int_{0}^{\chi_*}d\chi e^{-\tau} \frac{\partial \phi_L}{\partial \chi},
\ee
where $\tau$ is the optical depth defined as the line of sight integral of number density of free electrons $n_e$ as $\tau=\int_{\eta_i}^{\eta_0}n_e\sigma_T d\eta$.
Note that for $z>z_{\oplus}$, similar to the CMB lensing case, we have $\Phi_L\rightarrow \Psi$. Due to the cosmic variance effect, it is hard to find the signal in CMB alone. Accordingly, the ISW effect is extracted via the cross-correlation with the galaxy distribution.  The sophisticated idea is that the same potential which affects the CMB anisotropies is the same potential that governs the dynamics of the dark matter halos. Halos are the host of the galaxies.
Accordingly, the ISW signal is extracted via the ISW-galaxy cross-correlation function or its angular power spectrum $C_{\ell}^{gT}$ as \cite{Amendola2011}
\be
C_{\ell}^{gT}= \frac{2}{\pi}\int_0^{\infty} k^2 dk I_{\ell}^{\text{ISW}}(k)I^{g}_{\ell}(k) P^{\ddot{u}}_{m} (k),
\ee
where $P^{\ddot{u}}_{m}(k)$ is the matter power spectrum of \"u$\Lambda$CDM in the present time which must be transfered to the $\Lambda$CDM power spectrum in redshifts $z>z_{\oplus}$. Note that $I_{\ell}^{\text{ISW}}(k)$ and $I^{g}_{\ell}(k)$ are the kernels of ISW and galaxy distribution defined as
\be
I_{\ell}^{\text{ISW}}(k) = \frac{3H_0^2\Omega_m}{k^2}\int dz \frac{d}{dz} \left[ \frac{D ^{\ddot{u}}(z)(1+z)}{D ^{\ddot{u}}(z=0)\xi(z)} \right] j_{\ell}(k\chi(z)),
\ee
\be
I^{g}_{\ell}(k)= \int dz b(k,z) \frac{dN}{dz} \left[ \frac{D ^{\ddot{u}}(z)}{D ^{\ddot{u}}(z=0)\xi(z)} \right] j_{\ell}(k\chi(z))
\ee
where $ \frac{dN}{dz}$ is the distribution function of galaxies in redshift, which is used in the cross-correlation procedure and is normalized to unity $\int dz \frac{dN}{dz}=1$ . Note that $b(k,z)$ is the bias parameter of galaxies distribution to dark matter halos and $ j_{\ell}$ is the spherical Bessel function which depends on comoving distance and wave number.
In this work, we assume that the bias parameter is unity. Note that cosmology can change the bias parameter as well. This effect in ISW-galaxy cross-correlation is studied \cite{Khosravi:2015boa} . In this work, we assume that the bias is almost unchanged.
It is worth mentioning again that, although the Poisson equation and the relation of the lensing potential to $\Psi$ are changed in \"u$\Lambda$CDM,  the lensing potential is unchanged in transition redshift.
 In the next section, we present the methodology and observational data which we used in this study.


\section{Methodology and Observational Data}
\label{Sec4}
\begin{figure*}
\centering
\includegraphics[width=\columnwidth]{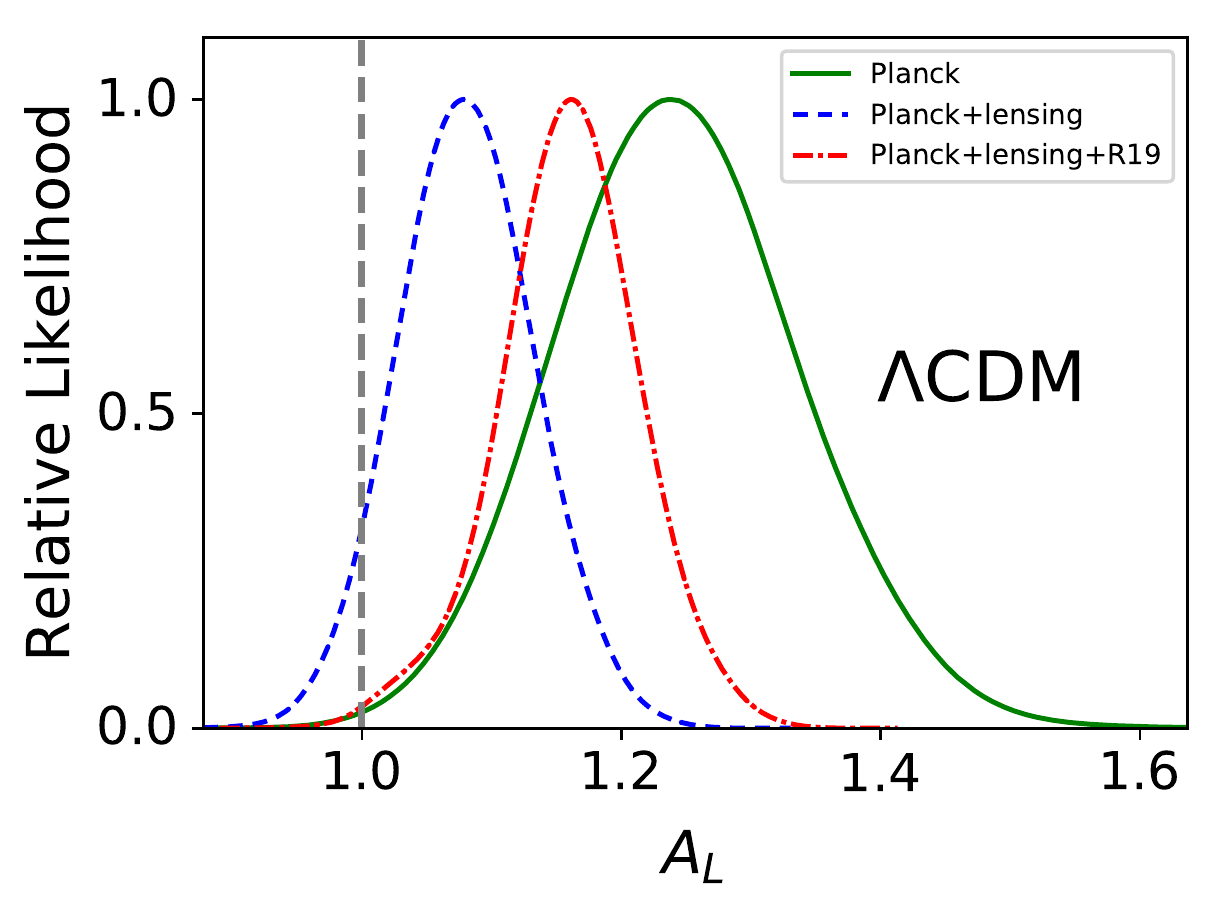}
\includegraphics[width=\columnwidth]{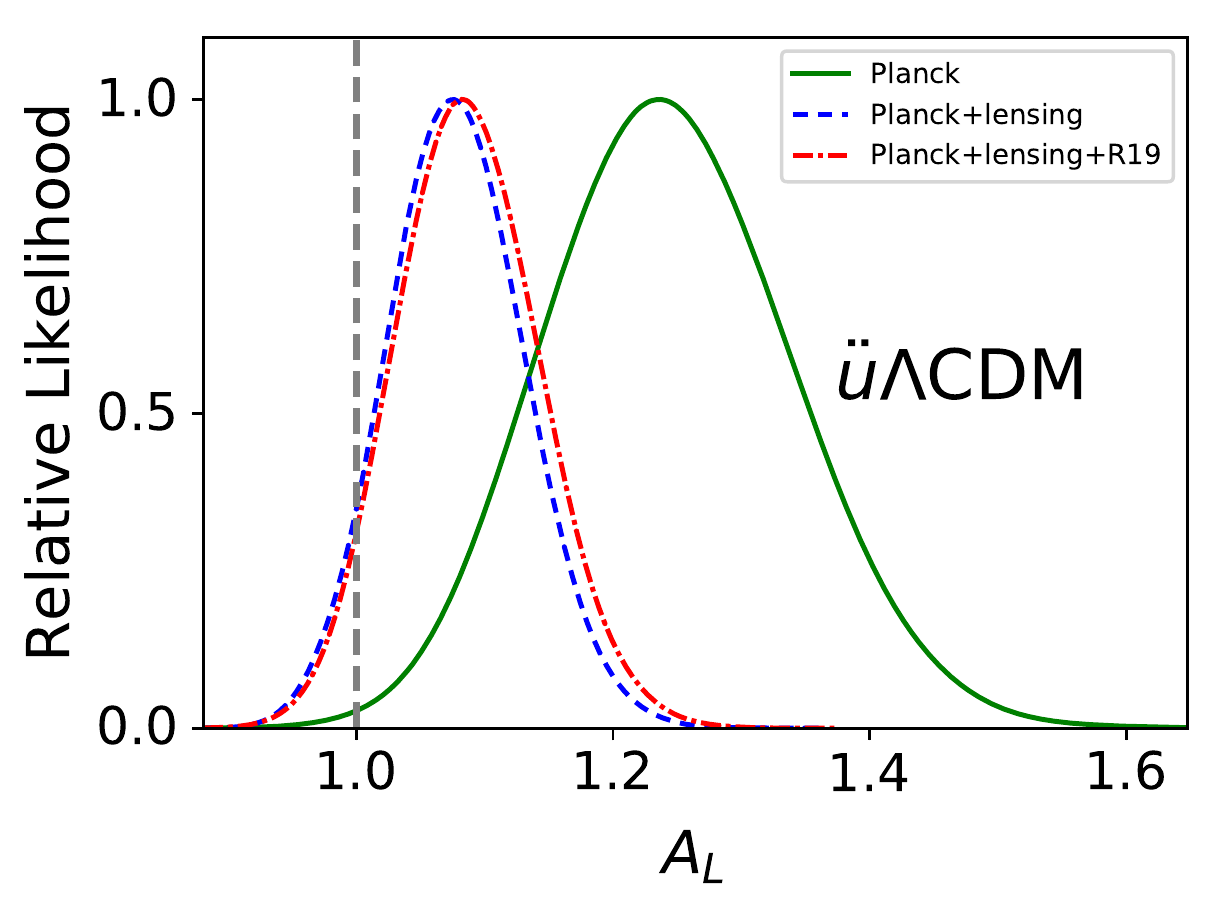}
\captionsetup{justification=raggedright,singlelinecheck=false}
\caption{The relative likelihood of the $A_L$ parameter is plotted for the standard $\Lambda$CDM and \"u$\Lambda$CDM models in left and right, respectively. Both models have $\sim2-3\sigma$ tension when we checked them against only the Planck dataset (green curves). When we add the lensing data (blue curves), both models are consistent with $A_L=1$ in their 1$\sigma$ regions. Adding R19 (red curves) restores the $2-3\sigma$ tension for the $\Lambda$CDM. CMB lensing and R19 datasets are not compatible in the $\Lambda$CDM. However \"u$\Lambda$CDM shows no inconsistency between lensing and R19 datasets. }\label{fig:AL}
\end{figure*}
\begin{figure*}
\centering
\includegraphics[width=\columnwidth]{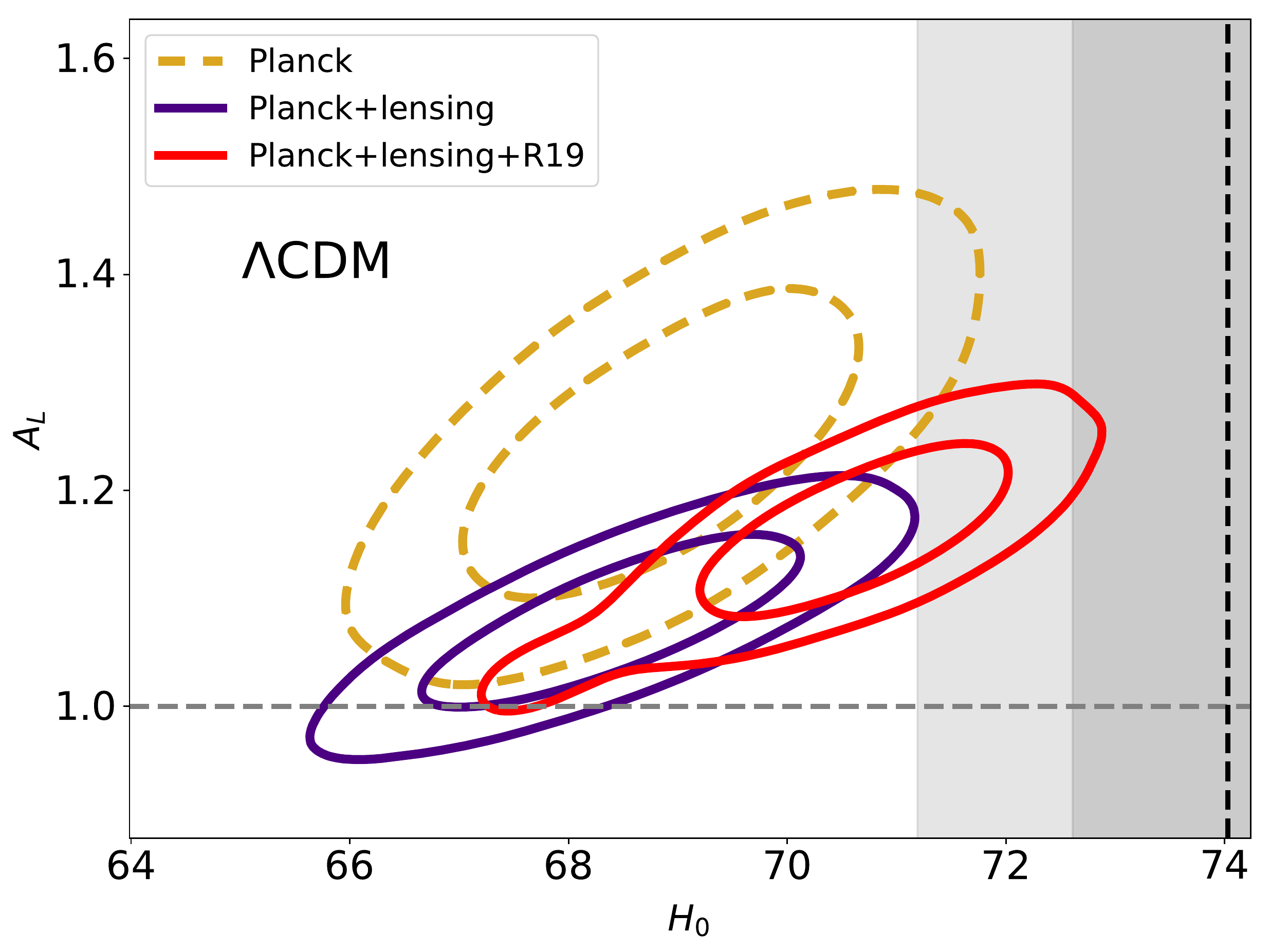}
\includegraphics[width=\columnwidth]{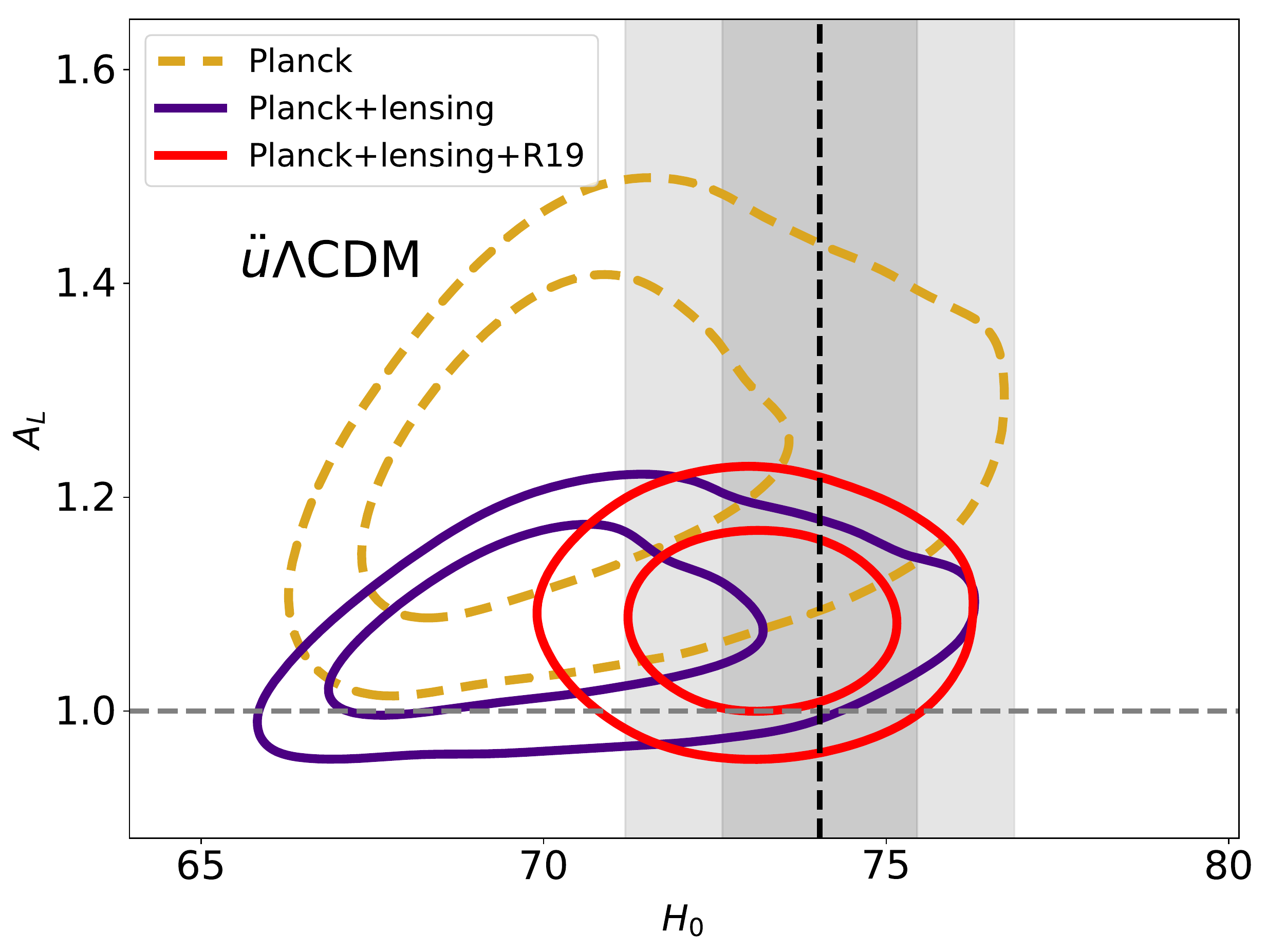}
\captionsetup{justification=raggedright,singlelinecheck=false}
\caption{We have plotted two-dimensional likelihood for $A_L$ vs $H_0$ for $\Lambda$CDM and \"u$\Lambda$CDM in the left and right, respectively. All the datasets' combinations of the $\Lambda$CDM model show a positive correlation between $A_L$ and $H_0$. That is why it is difficult for  $\Lambda$CDM to be consistent for both $A_L=1$ and R19 simultaneously. However, this correlation seems broken for the case of \"u$\Lambda$CDM}.\label{fig:AL-H0} 
\end{figure*}
\begin{figure}
\includegraphics[width=\columnwidth]{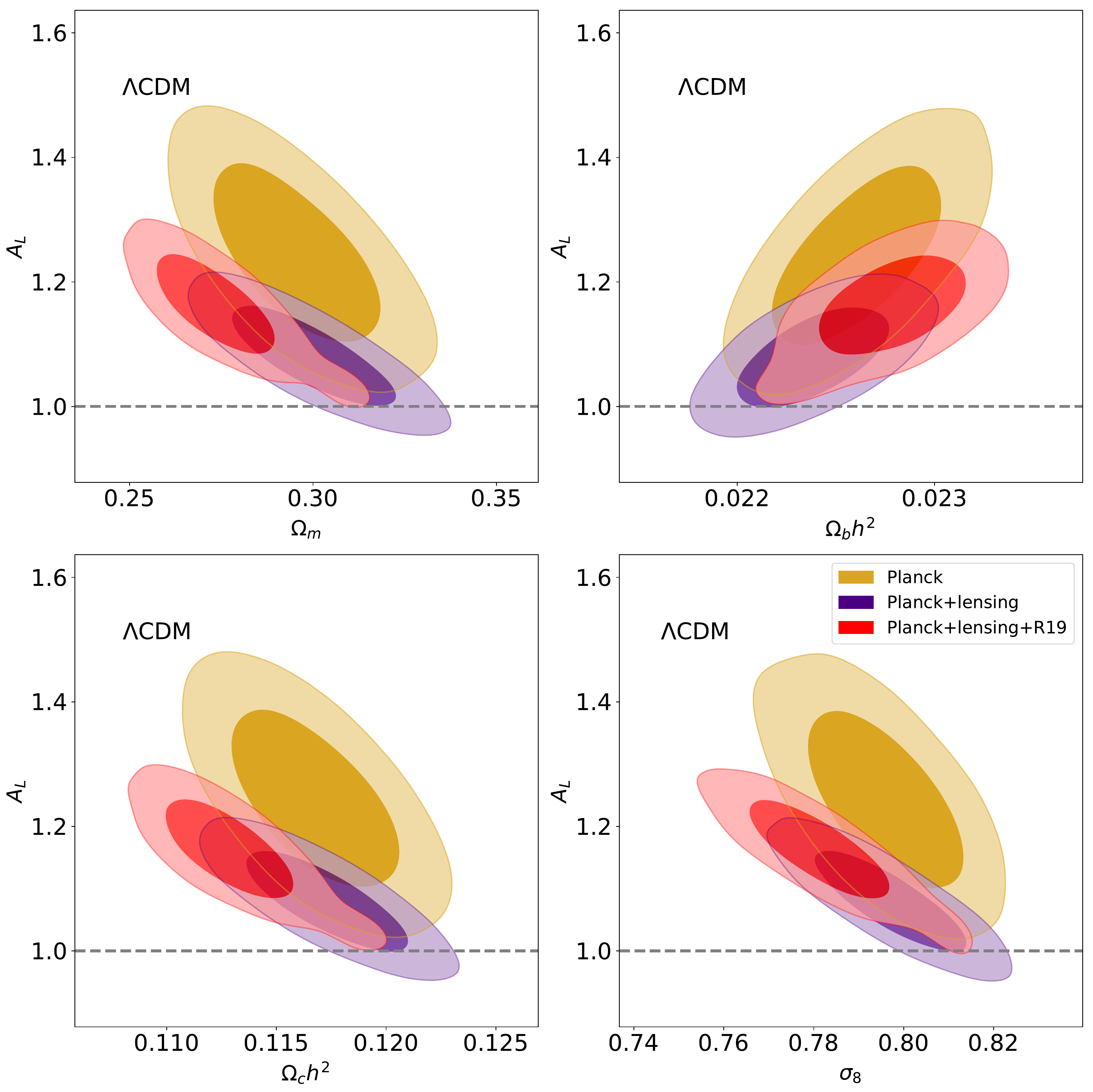}
\captionsetup{justification=raggedright,singlelinecheck=false}
\caption{The standard model's two-dimensional contour plots investigating the lensing amplitude versus matter density parameters of total matter, baryons, cold dark matter, and $\sigma_8$ are plotted.} \label{fig:2D-LCDM_AL}
\end{figure}
In this section, we use a combination of recent early universe and late-time measurements to constrain the \"u$\Lambda$CDM as follows:

\begin{itemize}
\item{ \bf Cosmic Microwave Background:}  We consider the CMB temperature, polarization, and lensing reconstruction angular power spectra as measured by the 2018 Planck legacy release ~\cite{Aghanim:2018eyx,Aghanim:2019ame}. We denote "Planck," which includes the CMB temperature and polarization data (TT, TE, EE+lowE; where the low-multipole polarization is obtained from the high-frequency instrument, HFI), and also we denote "Planck+lensing" which additionally includes the lensing reconstruction (TT, TE, EE+lowE+lensing).

\item {\bf Hubble Space Telescope (HST):}  We also use the recent estimation of the Hubble constant, $H_0 = 74.02 \pm 1.42$ at 68\% confidence level (C.L.) obtained from the Hubble Space Telescope \cite{Riess:2019cxk}. In this paper, we refer to this data as "R19".

\item {\bf ISW-galaxy cross-correlation:} For this observable we use  NRAO VLA Sky Survey (NVSS) \cite{Ho:2008bz} and  Wide-field Infrared Survey Explorer (WISE) dataset \cite{Wright:2010qw,Ferraro:2014msa}
\end{itemize}

In our cosmological analysis, we perform Markov chain Monte Carlo (MCMC) calculations with a modified version of \textsc{MGCosmoMC} publicly available code \cite{Hojjati:2011ix} and standard public code \textsc{CosmoMC} \cite{Lewis:2002ah}. We use a convergence criterion that obeys $R-1 < 0.01$, where the Gelman-Rubin $R$-statistics \cite{Gelman92} is the variance of chain means divided by the mean of chain variances.\\
We consider a seven parameters model with six of the $\Lambda$CDM model parameters plus lensing amplitude parameter $A_L$. Also, for the \"u$\Lambda$CDM model we consider an extra parameter, transition scale factor $(a_\oplus = (1+z_\oplus)^{-1})$. Assumption for priors of parameters are listed in Table~\ref{tabpriors}.
\begin{table}[tb]
\begin{tabular}{lll}
\hline
\hline
Parameter & Symbol & Prior\\
\hline
Cold dark matter density & $\Omega_{\mathrm c}h^2$ & $[0.001, 0.99]$\\
Baryon density & $\Omega_{\mathrm b}h^2$ & $[0.005, 0.1]$\\
Transition scale factor & $ a_\oplus$ & $[0.4, 1.0]$\\
Lensing amplitude & $ A_L$ & $[0.5, 3.0 ]$\\
Amplitude of scalar spectrum & $\ln{(10^{10} A_{\mathrm s})}$ & $[1.61, 3.91]$\\
Scalar spectral index & $n_{\mathrm s}$ & $[0.8, 1.2]$\\
Angular scale at decoupling & $100 \Theta_{\rm {MC}}$ & $ [0.5, 10.0]$\\
Optical depth & $\tau$ & $[0.01, 0.8]$ \\
Pivot scale $[{\rm{Mpc}}^{-1}]$ & $k_{\rm pivot}$ & 0.05 \\
\hline
\end{tabular}
\captionsetup{justification=raggedright,singlelinecheck=false}
\caption{Flat priors on the cosmological parameters varied in this paper.} \label{tabpriors}
\end{table}
In the next section, we will discuss the observational constraints on the model's free parameter based on lensing observations.


\section{Results: Parameter Estimation}
\label{Sec5}
In this section we examine the standard model and \"u$\Lambda$CDM, facing the CMB Planck data \cite{Aghanim:2018eyx} and CMB lensing data \cite{Aghanim:2018oex} and also the $H_0$ measurements by Riess {\it{et al.}}\cite{Riess:2019cxk}.
\begin{table*}[tb]
\begin{center}
\resizebox{\textwidth}{!}{
\begin{tabular}{ c |c c c ||c c c }
\hline
\hline
& $\Lambda$CDM & $\Lambda$CDM & $\Lambda$CDM & \"u$\Lambda$CDM & \"u$\Lambda$CDM & \"u$\Lambda$CDM \\ \hline
Parameters & Planck& Planck+ lensing & Planck +lensing+R19 & Planck & Planck+ lensing & Planck +lensing+R19 \\ \hline

$\Omega_m$ & $0.295\pm 0.015$ & $0.300^{+0.014}_{-0.015}$& $0.2744^{+0.0091}_{-0.012}$ & $0.280^{+0.022}_{-0.017}$ & $0.300^{+0.014}_{-0.015}$& $0.260^{+0.010}_{-0.011}$ \\

$H_0$ & $68.9\pm 1.2$ & $68.4\pm 1.1$ & $70.5^{+1.0}_{-0.85}$ & $70.7^{+1.4}_{-2.6} $ & $70.4^{+1.5}_{-2.6} $ & $73.2\pm 1.3 $ \\

$a_\oplus$ & $ -$ & $ -$ & $ -$ & $ > 0.687$ & $0.75^{+0.13}_{-0.17}$ & $ 0.607^{+0.031}_{-0.072}$ \\

$A_L$ & $1.244\pm 0.095$ & $1.081\pm 0.053$ & $1.159\pm 0.056$ & $1.244\pm 0.096$ & $1.078\pm 0.053$ &$1.087\pm 0.056$ \\

$S_8$ & $0.789\pm 0.030$ & $0.797\pm 0.029$ & $0.748^{+0.020}_{-0.026}$ & $0.826^{+0.038}_{-0.042}$ & $0.838^{+0.039}_{-0.044}$ & $0.854^{+0.041}_{-0.035}$ \\

\hline

$\chi^2_{\rm lensing} $ & $-$ & $10.0\,({\nu\rm{:}\,1.9})$ & $9.98\,({\nu\rm{:}\,2.3})$ & $-$ & $9.9\,({\nu\rm{:}\,2.0}) $ & $9.6\,({\nu\rm{:}\,2.2}) $ \\

$\chi^2_{\rm H074p03} $ & $-$ & $-$ & $6.5\,({\nu\rm{:}\,7.7}) $ & $-$ & $-$ & $1.2\,({\nu\rm{:}\,1.5}) $ \\

$\chi^2_{\rm CMB}$ & $624.2\,({\nu\rm{:}\,8.7})$ & $637.7\,({\nu\rm{:}\,7.0})$ & $641.2\,({\nu\rm{:}\,11.3})$ & $624.1\,({\nu\rm{:}\,7.3})$ & $637.6\,({\nu\rm{:}\,7.7})$ & $637.9\,({\nu\rm{:}\,8.0})$ \\

\hline
\hline

\end{tabular}
}
\end{center}
\captionsetup{justification=raggedright,singlelinecheck=false}
\caption{$68\%$ C.L. constraints for the \"u$\Lambda$CDM and $\Lambda$CDM models , for Planck, Planck+CMB lensing and Planck+CMB lensing+R19. Note that $ a_{\oplus} =(1+z_{\oplus})^{-1}$ and $\nu$ is degrees of freedom for $\chi^2$-test.}
\label{table_results}
\end{table*}
In the left panel of Fig. \ref{fig:AL}, we plot the likelihood of the $A_L$ parameter for the six parameter standard $\Lambda$CDM with CMB Planck data, which shows a tension for the Lensing amplitude. Adding the free parameter $A_L$ reconciles this tension. Interestingly adding the local SNeIa data worsens the situation. In contrast, the proposed \"u$\Lambda$CDM is compatible with both CMB lensing and the R19 distance indicator result.
In the right panel of Fig. \ref{fig:AL}, the red dash-dotted line shows the combination of Planck+lensing+R19 data for \"u$\Lambda$CDM, which is almost compatible with $A_L\sim1$. This means that the \"u$\Lambda$CDM has the same number of parameters as the standard model with $A_L$. \\
\begin{figure}
\includegraphics[width=\columnwidth]{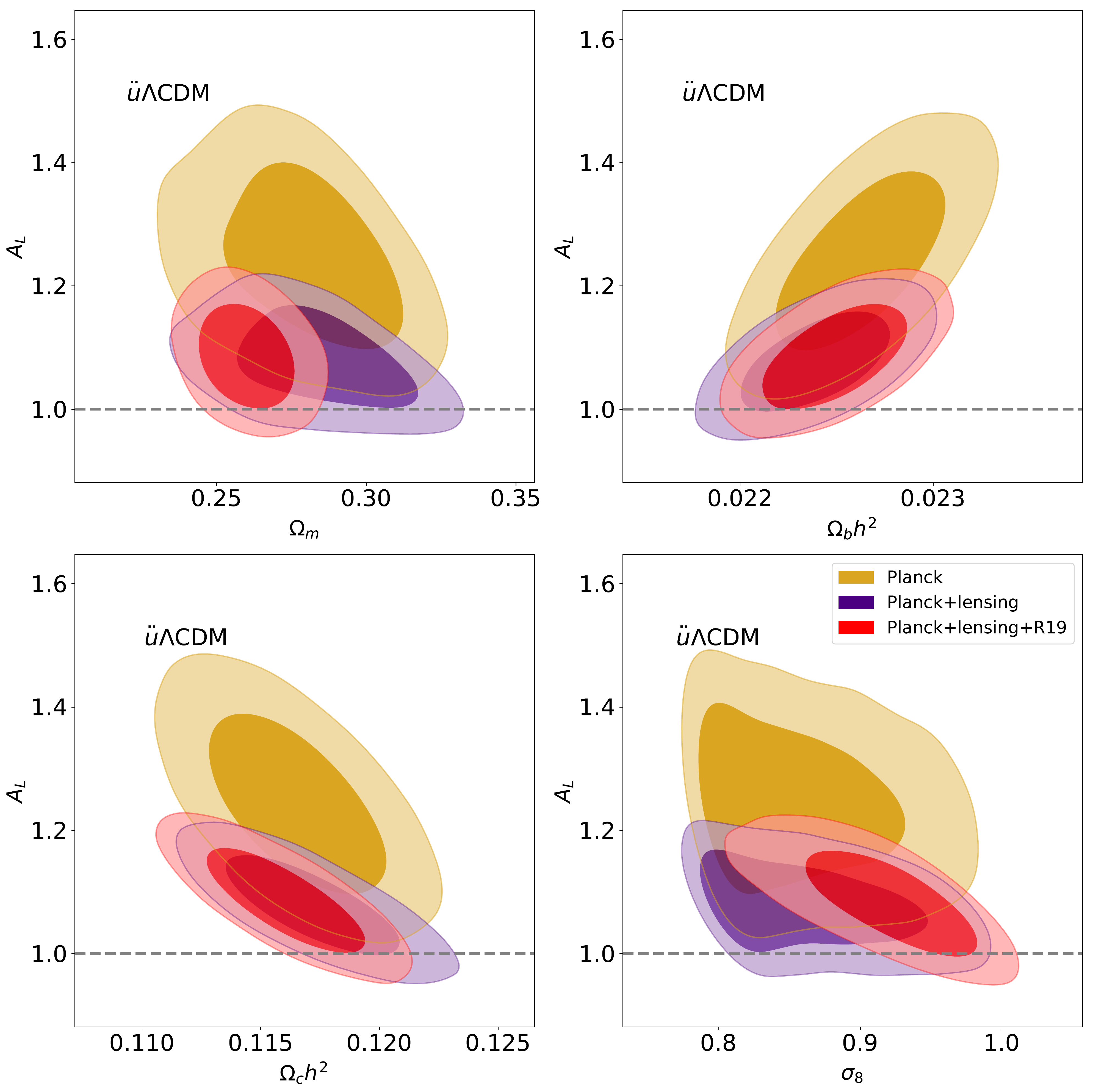}
\captionsetup{justification=raggedright,singlelinecheck=false}
\caption{Two-dimensional contour plots of \"u$\Lambda$CDM, investigating the lensing amplitude versus matter density parameters of total matter, baryons, cold dark matter, and $\sigma_8$, are plotted. } \label{fig:2D-ulcdm_AL}
\end{figure}
In Fig.\ref{fig:AL-H0}, we show the two-dimensional likelihood for $A_L$ vs $H_0$ for $\Lambda$CDM and \"u$\Lambda$CDM in the left and right panels, respectively. For all the datasets' combinations, the $\Lambda$CDM model shows a positive correlation between $A_L$ and $H_0$. That is why the $\Lambda$CDM cannot be consistent for both $A_L=1$ and R19 simultaneously. However, this correlation seems broken for the case of \"u$\Lambda$CDM. The modified gravity nature of \"u$\Lambda$CDM gives the opportunity to avoid the cycle of positive correlation of $A_L$ and $H_0$ and the negative correlation of $A_L$ and $\Omega_m$. \\
In Fig.\ref{fig:2D-LCDM_AL},  we plot the two-dimensional contour plots of the standard model, investigating the lensing amplitude versus matter density parameters of total matter, baryons, cold dark matter, and $\sigma_8$.\\
In Fig.\ref{fig:2D-ulcdm_AL}, we plot the confidence level of \"u$\Lambda$CDM for the lensing amplitude $A_L$ versus matter density parameters of total matter, baryons, cold dark matter, and $\sigma_8$. We find the anticipated correlations of $A_L$ with the matter density. It is worth mentioning that \"u$\Lambda$CDM shifts the $\sigma_8$ to the higher values, which will be a problem to compare this value with late-time observations. However, to further investigate this problem, we should look at the nonlinear structure formation in \"u$\Lambda$CDM to compare it with cluster count and weak lensing data in a nonlinear regime.
In Table \ref{table_results} we summarize all the results.
As we discussed in Sec. \ref{Sec3}, the ISW effect is the other secondary effect on CMB. We introduce the ISW to study the change of the gravitational potential in the dark energy-dominated era. In Fig. \ref{fig:iswcross}, we plot the ISW-galaxy cross-correlation function for both models of $\Lambda$CDM (black solid line) and \"u$\Lambda$CDM (blue dashed line). We compare the models with the observational data of NVSS and WISE. We show that the \"u$\Lambda$CDM is compatible with the ISW-galaxy cross-correlation data and it has an enhanced power in small angular scales due to change of the gravitational potential in transition redshift.
\begin{figure}
\includegraphics[width=\columnwidth]{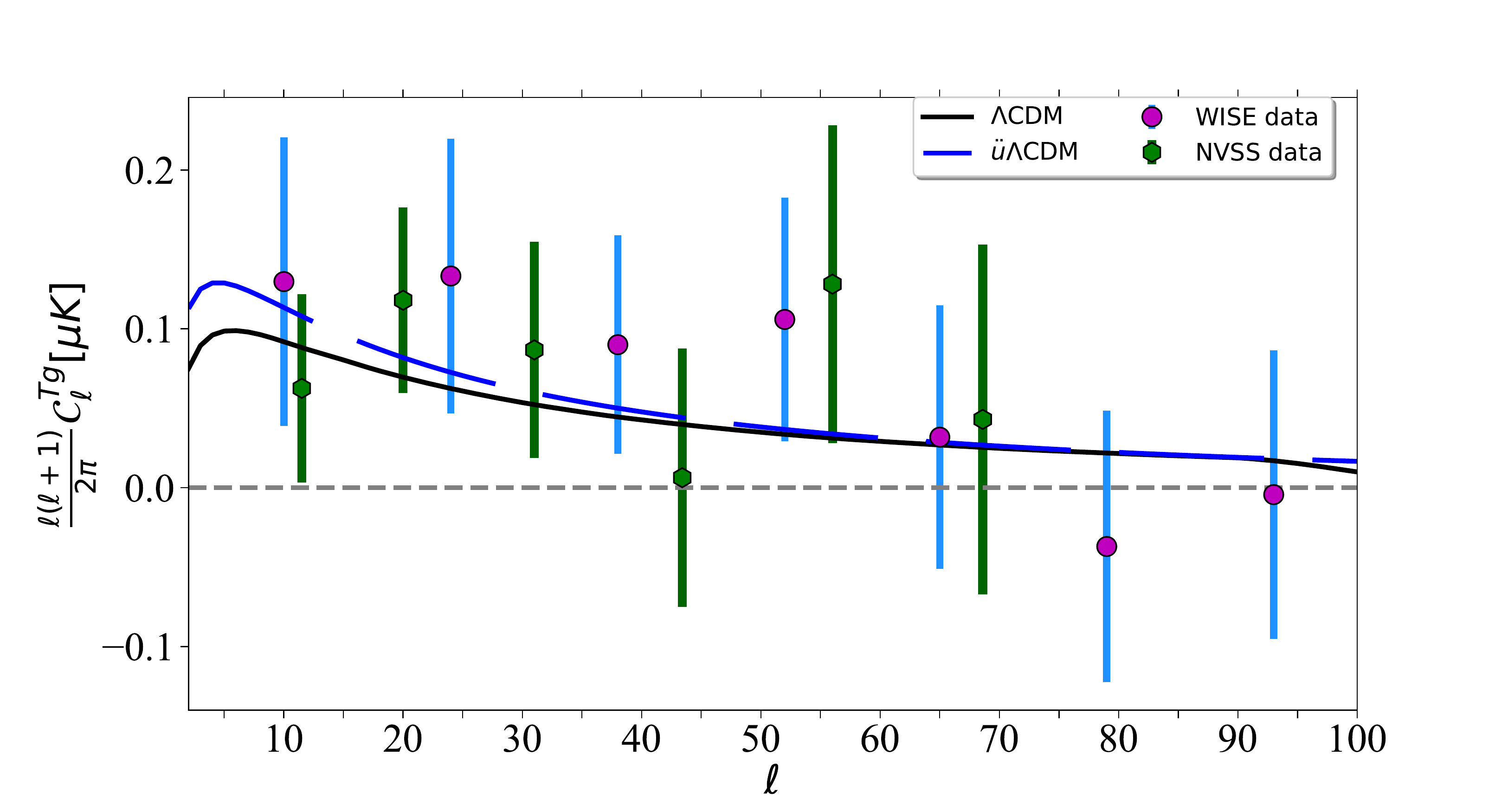}
\captionsetup{justification=raggedright,singlelinecheck=false}
\caption{ The ISW-galaxy angular cross power spectrum is plotted for $\Lambda$CDM (black solid line) and \"u$\Lambda$CDM (blue dashed line). The observational data are from the NVSS galaxy sample (hexagon points) and WISE (circle dots). } \label{fig:iswcross}
\end{figure}
As a final word to this section, we should note that there are arguments in literature on the validity of $H_0$ prior \cite{Benevento:2020fev,Camarena:2021jlr,Efstathiou:2021ocp}, as it is related to the absolute magnitude of SNe Ia and it is derived depending on the astrophysical model. This is not the essential concern in our work as we did not use SNe Ia data independently, which brings a twice usage of SNe Ia. Our transition in \"u$\Lambda$CDM occurred at higher redshifts $z\sim 0.6$, which is out of the scope of the priors studied in \cite{Efstathiou:2021ocp}. However, in a series of works, it is shown that the simultaneous use of the BAO and SNe Ia data prevent the late-time models to solve the $H_0$ tension \cite{Camarena:2021jlr,Efstathiou:2021ocp,Perivolaropoulos:2021jda,Banihashemi:2020wtb,Cai:2021weh}. In our analysis, we did not use these two datasets, as we are focusing on the secondary gravitational effects on the CMB. So, we should be conservative in our assertion admitting that it is unlikely that \"u$\Lambda$CDM can solve the $H_0$ tension fully once these additional datasets are used.

\section{Conclusion and Future remarks}
\label{Sec6}
In \cite{Khosravi:2017hfi} we have shown \"u$\Lambda$CDM can be a framework to study the $H_0$ tension. Here, we studied this model again to check if it can be a resolution for the CMB internal (mild) tensions. Interestingly, \"u$\Lambda$CDM which is first proposed to address the $H_0$ tension can solve the CMB anomalies. Our model does not see any discrepancies between low- and high-$\ell$'s power spectrum. The \"u$\Lambda$CDM has no tension with $A_L=1$ and R19 with just one additional free parameter in comparison to standard $\Lambda$CDM. {The $\Lambda$CDM with $A_L$ has an additional free parameter that can lessen $H_0$ tension but with a higher value for $A_L$. It means to address both $H_0$ and lensing tensions \"u$\Lambda$CDM is as good as $\Lambda$CDM+$A_L$ with the same number of free parameters.}
{Physically, higher $A_L$ means higher matter density, which is needed to solve the lensing anomaly. But in the \"u$\Lambda$CDM model, higher lensing is not because of higher matter density, but because of the stronger gravity. This is the property in the heart of the construction of the \"u$\Lambda$CDM, i.e., the way it is built.
We also investigate the ISW effect via the cross-correlation with the galaxy distribution. We check the consistency of \"u$\Lambda$CDM.

Finally, we should mention that stronger gravity may make the $\sigma_8$ tension worse. This issue needs more investigation in future works. Also, we should mention that as indicated beforehand as we did not use the SNe Ia and BAO, we should be conservative, indicating that it is unlikely that \"u$\Lambda$CDM can solve the $H_0$ tension fully once these additional datasets are used.
Also, an investigation on the recently raised issue of the magnitude versus Hubble tension is needed to study in detail for models like \"u$\Lambda$CDM, which has a higher redshift transition. Another way to pursue this model is by looking at the nonlinear structure formation. On the theory side, it will be interesting to look at it, if we can model the perturbation of \"u$\Lambda$CDM differently.} \\

\section*{ACKNOWLEDGMENTS} 
We thank the anonymous referee for his/her insightful comments and suggestion, which improved the manuscript extensively.
N.K. and S.B. are in debt to the Abdus Salam International Center of Theoretical Physics (ICTP) for the very kind hospitality. The main part of the idea of this work has been developed there. S.B. is partially supported by Abdus Salam International Center of Theoretical Physics (ICTP) under the junior associateship scheme during this work. This research is supported by Sharif University of Technology Office of Vice President for Research under Grant No. G960202.\\ \\

\end{document}